\documentclass{article}
\usepackage{amssymb,epsfig}

\def\dmatm{\Delta m_{@}^2}
\def\dmsun{\Delta m_{\odot}^2}
\def\tg2thsun{\tan^2\theta_{\odot}}
\def\eV{\mbox{\rm eV}}
\def\calN{{\cal N}}
\def\calO{{\cal O}}
\def\TT{^{\rm T}}
\def\diag{{\rm diag}}

\begin{document}
\begin{titlepage}
\title{
\vspace{2cm}
On the hierarchy of neutrino masses\thanks{
Presented by M.~Je\.zabek at 'Supersymmetry and Brane Worlds,' Fifth
European Meeting Planck 02, Kazimierz, Poland, May 25-29, 2002.}}
\author{M.~Je\.zabek$^{a,b}$
and
P.~Urban$^{a,c}$
\\ \\ \\ \small
$^a$ Institute of Nuclear Physics,
Kawiory 26a, PL-30055 Cracow, Poland
\\   \small 
 $^b$ Institute of Physics, University of 
      Silesia, \\[-5pt]  \small
     Uniwersytecka 4, PL-40007 Katowice, Poland\\ 
\small
$^c$ Institut f\"ur Theoretische Teilchenphysik,
Universit\"at Karlsruhe,\\  [-5pt]
\small
D-76128 Karlsruhe, Germany
}
\date{}
\maketitle
\thispagestyle{empty}
\vspace{-4.5truein}
\begin{flushright}
{hep-ph/0207018}\\
{TTP02-12}\\
{July 2002}
\end{flushright}
\vspace{4.0truein}
\begin{abstract}
We present a model of neutrino masses combining the seesaw mechanism and
strong Dirac mass hierarchy and at the same time exhibiting a
significantly reduced hierarchy at the level of active neutrino masses.
The heavy Majorana masses are assumed to be degenerate. The suppression
of the hierarchy is due to a symmetric and unitary operator $R$ whose
role is discussed. The model gives realistic mixing and mass
spectrum. The mixing of atmospheric neutrinos is attributed to the
charged lepton sector whereas the mixing of solar neutrinos is due to
the neutrino sector. Small $U_{e3}$ is a consequence of the model.
 The masses of the active neutrinos are given by
$\mu_3\approx\sqrt{\Delta m_{@}^2}$ and $\mu_1/\mu_2\approx
\tan^2\theta_\odot$.

\end{abstract}
\end{titlepage}
\section{Introduction}
In this talk the results on neutrino masses and mixings are presented
which were obtained in our recent publications \cite{MJ,MJPU}.

In view of the recent results from SNO \cite{SNO} and SuperKamiokande
\cite{SK-solar} and owing to developments in theory \cite{BPB}, see
also \cite{pdg00,bgg} and references therein,
there has emerged a unique solution to the problem of neutrino
oscillations. The only allowed solution is now LMA MSW, all others being excluded at the
$3\sigma$ level \cite{phenosol}. Thus we know the pattern of the oscillations of the
active neutrinos, that is those observed in experiment. Simultaneously it becomes more and more
clear that the oscillations of atmospheric neutrinos are due to
$\nu_{\mu}\rightleftarrows \nu_{\tau}$ transitions 
\cite{SK-atmo,K2K}. 
The third important piece of information is the CHOOZ limit \cite{CHOOZ} 
indicating that
the element $U_{e3}$ of the Maki-Nakagawa-Sakata (MNS) lepton mixing
matrix \cite{MNS} is small. 
 
The experimental data mentioned above can be described by a model based
on the seesaw mechanism \cite{seesaw} and a large hierarchy of the
Dirac masses of neutrinos. Such a model was considered in
 \cite{MJ,MJPU}. We present it here including some technical
details related to the derivation of the formulae.

The main idea of the present model is that 
a large hierarchy of the Dirac masses of neutrinos is possible even though the
hierarchy of masses indicated by experimental data is far less
pronounced than that expected from comparison with the quark or charged
lepton mass spectra. This nontrivial fact is interesting since it hints
at a possible similarity between the observed hierarchy of quark and
charged lepton masses
and that of neutrino Dirac masses. The disappearance of this hierarchy
at the level of the observable masses of the active neutrinos is caused
by the seesaw mechanism as well as by the algebraic structure of the low
energy effective mass operator $\cal N$ describing the masses of the
active neutrinos. The latter is due to a symmetric and unitary operator
$R$ acting in the flavor space and   related to the
unitary transformations of the right-handed neutrinos. This operator has
for the first time been considered in \cite{MJ}. It has been
pointed out in \cite{MJ,MJPU} that $R$ plays a crucial role in the
low energy physics of neutrinos. In fact, it affects the form of the
$U_{\rm MNS}$ mixing matrix. In the model we consider, we find a form of
the 
operator $R$ leading to a reduction of the underlying Dirac mass hierarchy
and thus producing realistic mass spectra. The resulting mixing matrix
is naturally exhibiting a small value of the $U_{e3}$ element. It also
follows from our model that the mass ratio of the two lighter neutrinos
is given by $\tan^2\theta_\odot$, $\theta_\odot$ denoting the solar
mixing angle.

\section{Mass hierarchy}
Our aim is to explain the observed mass spectrum of neutrinos starting
from a hierarchy of Dirac masses comparable with that of the
corresponding up quark masses. Let us begin with a look at data
and at the expectations from the simplest version of the  seesaw mechanism.
What is known are the squared mass differences affecting the oscillation
pattern of neutrinos. Denote the masses of the active neutrinos by
$\mu_1,\mu_2$ and $\mu_3$. Then we can define the ratio 
\begin{equation}
\rho = \frac{\dmsun}{\dmatm} = \frac{\mu_2^2-\mu_1^2}{\mu_3^2-\mu_2^2}
\label{eq:rho}
\end{equation}
of the solar to atmospheric mass splitting. The experimental value
of the parameter $\rho$ is
\begin{equation}
\rho_{exp} \approx \frac{5\cdot 10^{-5} \eV^2}{2.5 \cdot 10^{-3}
  \eV^2}=2\cdot 10^{-2}.
\label{rhoexp}
\end{equation}
Although this might well be called a hierarchy, we must compare it to the
predictions offered by the seesaw mechanism. Choose a reference frame
where the heavy Majorana mass matrix $M_R$ is diagonal and assume
furthermore that
it is proportional to the unit matrix. 
Of course, the hierarchy of the active neutrino masses can be
reduced by assuming a hierarchy in $M_R$, partly compensating for the
hierarchy originating from the Dirac masses, see e.g.\@ \cite{Stech,BW}.
However, we will not have
to abandon the simple assumption of degenerate right-handed Majorana
masses to destroy the hierarchy. So, we do not 
consider the more general case, although it is not hard to do so. Thus,
\begin{equation}\label{eq:MRequal1}
M_R=M\cdot {\mathbf 1}\quad .
\end{equation}
At the same time, the Dirac mass matrix for neutrinos is written as
\begin{equation}\label{eq:N}
N=U_R m^{(\nu)} U_L
\end{equation}
with
\begin{equation}\label{eq:mnu}
m^{(\nu)}={\rm diag}(m_1,m_2,m_3) \quad .
\end{equation}
Then the mass spectrum of the active neutrinos is given by the effective
operator $\cal N$ of dimension five:
\begin{equation}\label{eq:defN}
\calN = N^{\rm T}M_R^{-1}N=U_L^{\rm T}
m^{(\nu){\rm T}}U_R^{\rm T}M_R^{-1}U_R m^{(\nu)}U_L.
\end{equation}
With the simplifying assumption (\ref{eq:MRequal1}),
the mass spectrum obtained from the matrix
$\calN$ in eq.(\ref{eq:defN}) is  
seen to depend crucially on the following matrix $R$,
\begin{equation}\label{eq:defR}
R=U_R^{\rm T}U_R,
\end{equation}
which is symmetric and unitary. The matrix $U_R$ satisfying the equation
above for our final choice of $R$, see eq.(\ref{eq:R0}), can be found as described in Appendix A. The predictions of the
simplest seesaw 
model correspond to assuming that $R=1$. Then the resulting spectrum of
the active neutrino masses is
\begin{equation}\label{ActiveHierarchy}
\mu_1=\frac{m_1^2}{M} \ll \mu_2 = \frac{m_2^2}{M} \ll \mu_3 =
\frac{m_3^2}{M}\quad .
\end{equation}
Since we also require a hierarchy for the Dirac masses,
\begin{equation}
m_1 \ll m_2 \ll m_3,
\label{eq:hierarchical}
\end{equation}
it becomes evident that the active neutrino hierarchy is even stronger. 
 The ratio $\rho$ can now be estimated by letting the mass ratio
 $m_3/m_2$ be of the order of the corresponding mass ratios for other
 fundamental fermions, i.e. $m_b/m_s\sim 30$, $m_\tau/m_\mu \approx 17$
 or $m_t/m_c \sim 100$. We would obtain a quantity of the order of
 $10^{-8}$-$10^{-4}$, which is much less than the observed
 hierarchy. So, if we are to succeed in describing reality with a seesaw
 model, we must find some way of hiding this huge hierarchy.

Now we see that the operator $R$ cannot be a unit matrix. In fact, one
can easily convince oneself that its element $(R)_{33}$ must vanish in
order to prevent the Dirac mass hierarchy from showing up in the observable
mass ratio. 

Now consider what happens if the element $(R)_{23} = (R)_{32}$ is
non-vanishing. It turns out that the resulting
mass spectrum for the active neutrinos is acceptable from the phenomenological
point of view if $(R)_{23} = {\cal O}(1)$ is assumed. This spectrum 
corresponds to the case of the so-called inverted hierarchy. However,
the resulting structure of
the lepton mixing matrix does not resemble the experimentally observed
one \cite{MJPU}.

The only remaining case is $R_{33}=R_{23}=0$ which implies
\begin{equation}\label{eq:R0}
R= \left( 
\matrix{ 0 & 0 & \exp i\phi_1 \cr
         0 & \exp i\phi_2 & 0 \cr
         \exp i\phi_1 & 0 & 0 \cr }\right)
\end{equation}
The complex phase factors in eq.(\ref{eq:R0}) can be of crucial
importance for lepton number violating processes like neutrino-less
double beta decays. However, these phase factors do not affect our 
discussion which concentrates on neutrino oscillations. So, for the 
sake of simplicity, in the following considerations we take the same
form of $R$ as in \cite{MJ}:
\begin{equation}\label{eq:R}
R= \left( 
\matrix{ 0 & 0 & 1 \cr
         0 & 1 & 0 \cr
         1 & 0 & 0 \cr }\right)
\equiv P_{13} .
\end{equation}
It turns out that for strongly hierarchical Dirac masses eq.(\ref{eq:R}) 
is a necessary condition for a realistic mixing and mass spectrum. 
Therefore we assume that some symmetry underlying flavor
dynamics forces $U_R$ to fulfil eq.(\ref{eq:R}). 
The matrix $R$ can drastically reduce the hierarchy of the mass spectrum
for the active neutrinos. So, $R$ is observable, in principle at least, if
a large hierarchy of the Dirac masses is a common feature of all quarks
and leptons. In this sense $R$ is a physical object which is imprinted in 
low energy physical quantities, namely the masses of the active neutrinos.
Unlike the quark sector with its Cabibbo-Kobayashi-Maskawa mixing 
matrix \cite{CKM} the lepton sector has therefore two important
matrices in the flavor  
space. One is the lepton mixing matrix $U_{\rm MNS}$ \cite{MNS} 
which affects the form of the weak charged current. 
Another is the matrix $R$ defined in eq.(\ref{eq:defR}). 
$R$ affects the form of $U_{\rm MNS}$. Moreover, it is also reflected
in the low energy neutrino mass spectrum. In our phenomenological approach
we use the experimental input to fix the form of $R$. One may hope that
this is a first step towards an underlying theory of flavor. 

\section{Lepton mixing matrix}
In the previous Section we have arrived at a way of resolving the
problem of strong hierarchy of active neutrino masses. However, we must
show that the model describes correctly the mixing pattern. We now study
the mixing matrix $U_{\rm MNS}$. In our model, the mixing is due to both
charged leptons and neutrinos. The mass matrix for the charged leptons can be written as
\begin{equation}\label{eq:L}
L = V_R \; {\rm diag}(m_e,m_\mu,m_\tau)\; V_L \equiv V_R \;m^{(l)}\; V_L
\label{eq:1}
\end{equation}
The matrix $V_R$ multiplying $m^{(l)}$ from the left side can be made
equal to one by an appropriate redefinition of the right-handed charged
leptons. This has no observable consequences because at our low energies
only left-handed weak charged currents can be studied. The corresponding
Dirac mass matrix for the neutrinos is given in eq.(\ref{eq:N}).
Let $\calO$ be a unitary matrix such that 
\begin{equation}
\calO\TT \calN \calO = \diag(\mu_1,\mu_2,\mu_3) .
\label{eq:calO}
\end{equation}
Note that in eq.(\ref{eq:calO}) the diagonalization is done by
multiplying by the transposed matrix $\calO\TT$, rather than the
Hermitian conjugate $\calO^\dagger$,  from the left. We explain in
Appendix B how to 
perform such a diagonalization.
Eq.(\ref{eq:L}) implies that $M_L^2=L^\dagger L$ is diagonalized by
$V_L$, i.e. 
\begin{equation}
  \label{eq:Vdiagonalization}
  V_L M^2_L V_L^\dagger =
\diag(m_e^2,m_\mu^2,m_\tau^2).
\end{equation}
Then from eqs.(\ref{eq:defN},\ref{eq:defR})
one derives
\begin{equation}\label{UMNSVLO}
U_{\rm MNS}=V_L \calO = V_L U_L^{-1} \calO'
\end{equation}
where the unitary matrix $\calO'$ is such that
\begin{equation}
\frac{1}{M} {\calO'} \TT {m^{(\nu)}}\TT\; R\;  m^{(\nu)}
\calO'=\diag(\mu_1,\mu_2,\mu_3)
\label{eq:18}
\end{equation}
with $R = P_{13}$, cf. eq.(\ref{eq:R}).

We narrow our search for a realistic neutrino mass matrix by
concentrating on the
following structure of the MNS matrix, which is known to successfully describe
data,
\begin{equation}\label{UMNSboxes}
U_{\rm MNS}=\underbrace{
\left(\matrix{
 \matrix{1} & \matrix{ 0\quad & 0} \cr
 \matrix{0 \cr 0} & 
\begin{picture}(50,20)(0,0)
\put(0,-10){\line(0,1){30}}
\put(0,20){\line(1,0){50}}
\put(0,20){\line(5,-3){50}}
\put(0,14){\line(5,-3){40}}
\put(0,8){\line(5,-3){30}}
\put(0,2){\line(5,-3){20}}
\put(0,-4){\line(5,-3){10}}
\put(10,20){\line(5,-3){40}}
\put(20,20){\line(5,-3){30}}
\put(30,20){\line(5,-3){20}}
\put(40,20){\line(5,-3){10}}
\end{picture}
\cr\ }\right)
}_{\rm  charged\ lepton\ sector}
\underbrace{
  \left(\matrix{ 
\begin{picture}(50,20)(0,0)
\put(0,-15){\line(1,0){50}}
\put(50,-15){\line(0,1){30}}
\put(0,15){\line(5,-3){50}}
\put(0,9){\line(5,-3){40}}
\put(0,3){\line(5,-3){30}}
\put(0,-3){\line(5,-3){20}}
\put(0,-9){\line(5,-3){10}}
\put(10,15){\line(5,-3){40}}
\put(20,15){\line(5,-3){30}}
\put(30,15){\line(5,-3){20}}
\put(40,15){\line(5,-3){10}}
\end{picture}
& \matrix{0 \cr 0} \cr
\cr
\matrix{0\quad & 0} & 1}\right)
}_{\rm  neutrino\  sector}\quad .
\end{equation}
Since we choose a simple form of $U_L=1$, eq.(\ref{UMNSVLO}) implies
that the second matrix on the right hand side of eq.(\ref{UMNSboxes}) is
equal to the matrix $\cal{O'}$ diagonalizing the light neutrino Majorana
mass matrix. Such a form of this matrix can be obtained even if the mass
matrix $\cal{N}$ is not of the block diagonal form suggested by
eq.(\ref{UMNSboxes}).
In fact, assuming a diagonal
 Dirac mass matrix as in eq.(\ref{eq:mnu}), one obtains the light Majorana
 mass matrix 
 \begin{equation}
   \label{eq:lightN}
\calN=\mu \left(\matrix{0 & 0 & r \cr 0 & 1 & 0 \cr r & 0 & 0}\right),
 \end{equation}
where  
\begin{equation}
  \label{eq:rmu}
   r = m_1 m_3/ m_2^2, \quad  \mu = m_2^2/M . 
\end{equation}
  The way to sidestep this
difficulty is to exchange the eigenvectors of the mass matrix with a
permutation $P_{23}$,
\begin{equation}
  \label{eq:P23}
  P_{23}=\left(\matrix{1&0&0\cr 0&0&1 \cr 0&1&0}\right),
\end{equation}
 see \cite{MJ,MJPU} for details. 
Then $\cal{N}$ in eq.(\ref{eq:lightN}) is diagonalized by the matrix
\begin{equation}
\calO'=P_{23}U_{12}\left(\pi/4\right),
\end{equation}
where
\begin{equation}\label{U12}
U_{12}\left( \alpha \right)=\left( 
\matrix{ i\cos\alpha   &   \sin\alpha & 0 \cr
        -i\sin\alpha   &   \cos\alpha & 0 \cr
         0                    &  0                     & 1}
\right).
\end{equation}

This way, the matrix $P_{23}$ will appear sandwiched between the two matrices
in eq.(\ref{UMNSboxes}). In principle, such an insertion could destroy
the structure of the mixing matrix. However, due to the particular form
of the charged lepton matrix, nothing wrong happens since the effect of
$P_{23}$ acting to the left is that of exchanging the second and third
column. But this may be seen to correspond to an innocuous relabelling
of flavors, made irrelevant especially for the model of the charged
lepton matrix we are using,
\begin{equation}
  \label{eq:VL}
  V_L=O_{23}(\pm \pi/4),
\end{equation}
the right hand side meaning the rotation about the first axis by the
angle of $\pm\pi/4$ \cite{MJPU}. This form of $V_L$ has been considered
in many published models \cite{AFreview}, in particular in the models
based on the so-called lopsided form of the charged lepton mass matrix \cite{ABB}.

\section{A realistic model}
The construction shown above lets us get rid of the factor of $m_3^2/M$ in 
the active neutrino spectrum, cf.\@ eq.(\ref{ActiveHierarchy}), but the mass
splitting ratio is now zero due to the twofold degeneracy of the lighter
states, so the hierarchy problem appears to have actually been
aggravated. On the other hand,
the mixing pattern corresponds to the so-called bimaximal mixing
\cite{bi-maxim} which
is not acceptable for the solar neutrinos \cite{phenosol}.
Both problems can be cured by an appropriate perturbation of the Dirac matrix,
whose form was found in \cite{MJPU}. The resulting low energy
neutrino mass matrix is
\begin{equation}\label{P23MP23}
{\cal M} = P_{23} U_L^* {\calN} U_L^{-1} P_{23}
=\mu\left(\matrix{0&r&0\cr r& 2ar&0\cr 0&0&1}\right), 
\end{equation}
where $ a^{-1}=\tan 2\theta_\odot $.
For $a>0$ the matrix ${\cal M}$ in (\ref{P23MP23}) is diagonalized by
\begin{equation}
U_{12}\TT(\theta_\odot) {\cal M} U_{12}(\theta_\odot) =\diag\left(\mu_1,\mu_2,\mu_3\right).
\label{eq:diagM}
\end{equation}
Realistic spectra and mixing are obtained for the following range of the
parameters $a$ and $r$ :
\begin{equation}\label{eq:rangeofar}
 0.35 \le a \le 0.75\quad {\rm\ and\ }\quad  0.05 \le r \le 0.25 
\end{equation}
with the best fits corresponding to $a$ between 0.46 and 0.57
and $r$ between 0.09 and 0.10.
It is interesting that the value of $r\approx 0.08$ is obtained if the
Dirac masses of neutrinos are assumed to be proportional to the
corresponding masses of the charged leptons, see the footnote after
eq.(42) in \cite{MJPU}.
The lepton mixing matrix becomes
\begin{equation}\label{UMNSVUPU}
U_{\rm MNS}=V_L U_L^{-1} P_{23} U_{12}\left( \theta_\odot \right),
\end{equation}
and our model leads to the following mass spectrum, see \cite{MJPU}:
\begin{eqnarray}
\mu_1 &\approx& \sqrt{\dmsun}\; \tan^2\theta_\odot\;/ \;
\sqrt{1 - \tan^4\theta_\odot} \label{eq:41}\\
\mu_2 &\approx& \sqrt{\dmsun} \;/ \; \sqrt{1 - \tan^4\theta_\odot} 
\label{eq:42}\\
\mu_3 &\approx& \sqrt{\dmatm} \label{eq:43}.
\end{eqnarray}

 From the presented model we can derive the lightest neutrino mass. One obtains
 about $3$ meV at $\tan\theta_\odot^2 \approx 0.4$.
This mass range can be probed by the
10t version of the GENIUS project \cite{GENIUS}. Finally,
let us note that the mass scale of the Majorana
masses is between $10^{10}$ and $10^{11}$ GeV if $m_2 \sim m_c$ 
is assumed. It has been pointed out in \cite{jez01} that 
this is exactly the range of Majorana masses which 
may be important for baryogenesis; 
see \cite{buchmuller} and references therein.

\section{Concluding remarks}
The observed neutrino mass splitting ratio exhibits little hierarchy
compared to the expectations from a simple seesaw model and the
assumption of a Dirac mass hierarchy of the order typical for
quarks. Nevertheless, we 
have shown that seesaw models exist which succeed in suppressing the
underlying strong Dirac mass hierarchy leaving only the weak hierarchy of
effective light Majorana masses of active neutrinos. This reduction of
hierarchy is caused by the symmetric unitary operator $R$, whose effect
is to modify both the low energy mass spectrum and the lepton mixing
matrix $U_{\rm MNS}$. Realistic mixing pattern and masses are obtained with
the form of $R$ proposed in \cite{MJ} after introducing a proper
perturbation of the diagonal Dirac mass matrix \cite{MJPU}. The resulting
model predicts a relation between the two lighter active neutrino
masses,
$\mu_1/\mu_2\approx \tan^2\theta_\odot$, which is stable under small
perturbations. Furthermore, the mass of the heaviest neutrino is related 
to the mass scale $\sqrt{\dmatm}$ governing the oscillations of atmospheric 
neutrinos. A small $U_{e3}$ follows naturally from the model. The mass of the
lightest neutrino is predicted to be about $3$ meV and can be tested by the
10t GENIUS detector if Majorana 
phases are not too small and there are no strong cancellations between 
contributions to the mass parameter $\left< m_{\nu_e} \right>$.

\section{Acknowledgments}

This talk was presented during a special session on the occasion of
the 60th birthday of Stefan Pokorski. We would like to express our
admiration for his works in physics and wish him many further
successes in the future.

This work was done during our stay in the Institut f. Theoretische
Teilchenphysik, Universit\"at Karlsruhe (TH).  
We would like to thank the Alexander-von-Humboldt Foundation for 
grants which made this possible. A warm atmosphere in TTP is gratefully
acknowledged. 

Work supported in part by the European Community's Human Potential
Programme under contract HPRN-CT-2000-00149 Physics at Colliders,
and by the KBN grant 5P03B09320.

\appendix

\section{Solution for $U_R$}
The unitary matrix $U_R$ defined in eq.(\ref{eq:N}) is to satisfy the relation
\begin{equation}
  \label{eq:URTUR}
  U_R^{\rm T}U_R=\left(\matrix{0&0&\exp i\phi_1 \cr 0&\exp i\phi_2&0 \cr
      \exp i\phi_1&0&0}\right).
\end{equation}
Note that if
\begin{equation}
  \label{eq:URprim}
  U_R'^{\rm T}U_R'=\left( 
\matrix{ 0 & 0 & 1 \cr
         0 & 1 & 0 \cr
         1 & 0 & 0 \cr }\right)
\end{equation}
and 
\begin{equation}
  \label{eq:URbyURprim}
  U_R=U_R'\left(\matrix{\exp \frac{i\phi_1}{2} &0&0\cr 0 & \exp
      \frac{i\phi_2}{2}&0\cr 0&0&\exp \frac{i\phi_1}{2}}\right),
\end{equation}
then the condition (\ref{eq:URTUR}) is satisfied. It is thus enough to
find the matrix $U_R'$ fulfilling eq.(\ref{eq:URprim}), which can be
represented as
\begin{equation}
  \label{eq:URprimP13URprimstar}
  U_R'^{\rm T}=\left( 
\matrix{ 0 & 0 & 1 \cr
         0 & 1 & 0 \cr
         1 & 0 & 0 \cr }\right)U_R'^{*{\rm T}}\quad .
\end{equation}
Denoting the elements of $U_R'^{\rm T}$ as
\begin{equation}
  \label{eq:URelem}
  (U_R'^{\rm T})_{ij}=a_{ij}
\end{equation}
one can write equations for the $a_{ij}$ following from eq.(\ref{eq:URprimP13URprimstar}):
\begin{equation}
  \label{eq:row13}
  a_{3i}=a_{1i}^*, \quad  a_{2i}=a_{2i}^*, \quad i=1\dots 3
\end{equation}
Another set of relations follows from the unitarity of $U_R'$ and can
conveniently be written in terms of the real vectors $\mathbf{u,v,w}$ defined as
\begin{equation}
  \label{eq:uvw}
  \mathbf{u}\equiv(a_{21},a_{22},a_{23}),\quad \mathbf{v}+i\mathbf{w}\equiv(a_{11},a_{12},a_{13}).
\end{equation}
Then,
\begin{equation}
  \label{eq:scalaruvw}
  \mathbf{u}\cdot\mathbf{v}=\mathbf{u}\cdot\mathbf{w}=\mathbf{v}\cdot\mathbf{w}=0,
\end{equation}
\begin{equation}
  \label{eq:scalarsame}
  \mathbf{u}^2=1, \quad \mathbf{v}^2=\mathbf{w}^2=\frac{1}{2}.
\end{equation}
Obviously, the conditions (\ref{eq:scalaruvw},\ref{eq:scalarsame}) do
not change under a rotation of the system:
\begin{equation}
  \label{eq:rotation}
  \left(\matrix{\mathbf{v}+i\mathbf{w}\cr
      \mathbf{u}\cr\mathbf{v}-i\mathbf{w}}\right)\longrightarrow
\left(\matrix{\mathbf{v}+i\mathbf{w}\cr
      \mathbf{u}\cr\mathbf{v}-i\mathbf{w}}\right)\cal{O},
\end{equation}
where $\cal{O}$ is an arbitrary orthogonal matrix. We can therefore
choose
\begin{equation}
  \label{eq:chooseuvw}
  \mathbf{u}=(0,1,0),\quad \mathbf{v}=\frac{1}{\sqrt{2}}(1,0,0),\quad 
  \mathbf{w}=\frac{1}{\sqrt{2}}(0,0,1)
\end{equation}
to get
\begin{equation}
  \label{eq:UR0}
  U_R'=\left(\matrix{\frac{1}{\sqrt{2}}&0&\frac{1}{\sqrt{2}}\cr 0&1&0\cr
                     \frac{i}{\sqrt{2}}&0&\frac{-i}{\sqrt{2}}}\right).
\end{equation}
If we take the rotation matrix
\begin{equation}
  \label{eq:rotateUR}
  \cal{O}=\left(\matrix{-\frac{1}{\sqrt{6}}&\frac{1}{\sqrt{3}}&\frac{1}{\sqrt{2}}\cr
 \sqrt{\frac{2}{3}}&\frac{1}{\sqrt{3}}&0\cr
 -\frac{1}{\sqrt{6}}&\frac{1}{\sqrt{3}}&-\frac{1}{\sqrt{2}}}\right),
\end{equation}
we obtain another solution to eq.(\ref{eq:URprim}),
\begin{equation}
  \label{eq:UR3}
  {\cal{O}} U_R'=\frac{1}{\sqrt{3}}\left(\matrix{\omega & 1 &\omega^*\cr
  1&1&1\cr \omega^* & 1 & \omega}\right),
\end{equation}
where $\omega=\exp \frac{2\pi i}{3}$.

\section{Diagonalization with transposed matrices}
When diagonalizing the neutrino mass matrix, one must do it by multiplying
a unitary matrix $V$ to the right and its {\em transpose}, $V^{\rm T}$,
rather than the Hermitian conjugate, to the left. That is, we are faced
with the problem of finding a unitary matrix $V$ such that a given
symmetrical matrix $M$ is diagonalized:
\begin{equation}\label{eq:diag}
V^{\rm T}M V = {\rm diag}(\lambda_1,\dots,\lambda_n).
\end{equation} 
Since the eigenvalues are to be interpreted as masses, we
furthermore require $\lambda_i \in \mathbf{R}$ and $\lambda_i \ge 0$.
In general, this problem is different from the usual procedure of
diagonalization. In this appendix we show that the matrix $V$ satisfying
eq.(\ref{eq:diag}) has the form
\begin{equation}\label{eq:Vsolve}
V=\left(\mathbf{v}_1\;\mathbf{v}_2\;\dots\;\mathbf{v}_n \right)=
 \left(\mathbf{u}_1\;\mathbf{u}_2\;\dots\;\mathbf{u}_n \right)-i
 \left(\mathbf{w}_1\;\mathbf{w}_2\;\dots\;\mathbf{w}_n\right)
\end{equation}
where $\mathbf{u}_i,\mathbf{v}_i,\mathbf{w}_i$ are column vectors, e.g.
\begin{equation}
  \label{eq:vcomponents}
  \mathbf{v}_i=\left(\matrix{v_{i,1}\cr\vdots\cr v_{i,n}}\right),
\end{equation}
and they satisfy the equation
\begin{equation}\label{eq:Mbig}
\mathcal{M}\left(\matrix{\mathbf{u}_i \cr \mathbf{w}_i }\right)\equiv\left(\matrix{M_R & M_I \cr M_I & -M_R }\right) 
\left(\matrix{\mathbf{u}_i \cr \mathbf{w}_i }\right) = \lambda_i 
\left(\matrix{\mathbf{u}_i \cr \mathbf{w}_i }\right) {\rm \ for\
  }\lambda_1,\dots,\lambda_n \ge 0.
\end{equation}
In the formula above, $M_R$ and $M_I$ are the real and imaginary parts
of the matrix $M$, respectively:
\begin{equation}
  \label{eq:MRMI}
  M=M_R+i M_I
\end{equation}
 Note that if 
\begin{equation}
  \left(\matrix{\mathbf{u}_i\cr
    \mathbf{w}_i}\right)
\end{equation}
is the eigenvector with the eigenvalue $\lambda_i$ then 
\begin{equation}
  \left(\matrix{\mathbf{w}_i\cr
    -\mathbf{u}_i}\right)
\end{equation}
is the eigenvector with the eigenvalue $-\lambda_i$.
The unitarity of $V$ requires the orthogonality relations for the vectors
$\mathbf{u}_i,\mathbf{w}_i$:
\begin{equation}
  \label{eq:VdagV}
  \mathbf{v}_i^\dagger \mathbf{v}_j = \mathbf{u}_i^{\rm T}\mathbf{u}_j + \mathbf{w}_i^{\rm T} \mathbf{w}_j = (\mathbf{u}_i, \mathbf{w}_i)^{\rm
    T} \left(\matrix{\mathbf{u}_j\cr \mathbf{w}_j}\right)=\delta_{ij}. 
\end{equation}
The conditions (\ref{eq:VdagV}) are fulfilled due to the fact that the
$2n \times 2n$ matrix $\mathcal{M}$ appearing in eq.(\ref{eq:Mbig}) is symmetric and
real, so it has orthogonal eigenvectors, which can be normalized.

To prove eq.(\ref{eq:Vsolve}), solve the equation
\begin{equation}
  \label{eq:MV}
  MV=V^*\Lambda, {\rm \ where\ } \Lambda={\rm
    diag}(\lambda_1,\lambda_2,\dots,\lambda_n).
\end{equation}
Writing 
\begin{equation}
  \label{eq:VRVI}
  V=V_R-iV_I,
\end{equation}
where
\begin{equation}
  \label{eq:VRVIdef}
  V_R=\left( \mathbf{u}_1 \; \mathbf{u}_2\;\dots\;\mathbf{u}_n \right),
  V_I=\left( \mathbf{w}_1\;\mathbf{w}_2\;\dots\;\mathbf{w}_n \right),
\end{equation}
we obtain $ (M_R+iM_I)(V_R-iV_I)=(V_R+iV_I)\Lambda$ and so
\begin{eqnarray}
  M_R V_R + M_I V_I & = & V_R \Lambda \\
  M_I V_R - M_R V_I & = & V_I \Lambda.
\end{eqnarray}
This system of equations can be rewritten in terms of the matrix
$\mathcal{M}$
\begin{equation}
  \label{eq:Msystem}
  \mathcal{M}\left(\matrix{\mathbf{u}_i \cr \mathbf{w}_i}\right)
=\lambda_i \left(\matrix{\mathbf{u}_i \cr \mathbf{w}_i}\right),\quad i=1\dots n.
\end{equation}
Therefore the problem is reduced to finding the eigenvalues and
eigenvectors of the matrix $\mathcal{M}$. There are $2n$ of them and
the eigenvalues are of the form $\pm \lambda_1, \pm \lambda_2, \dots \pm
\lambda_n$. Of those, we choose the $n$ non-negative ones.

\end{document}